\documentclass{article}
\usepackage[T1]{fontenc} 
\usepackage[utf8]{inputenc} 
\usepackage{ismir,amsmath,cite,url, amssymb}
\usepackage{graphicx}
\usepackage{color}
\usepackage{comment}
\usepackage{lineno}
\usepackage{siunitx}
\pdfoutput=1
\nolinenumbers

\newcommand{\Rnonlin}{$R_\mathrm{nonlin}$}

\newcommand{\SDRinp}{$\mathrm{SDR_{inp}}$}

\title{Distortion Audio Effects: \\Learning How to Recover the Clean Signal}

\newcommand\authorhspace{0.5cm}
\multauthor
{Johannes Imort$^\natural$ \hspace{\authorhspace} Giorgio Fabbro$^\sharp$ \hspace{\authorhspace} Marco A. Martínez-Ramírez$^\flat$} { \bfseries{Stefan Uhlich$^\sharp$ \hspace{\authorhspace} Yuichiro Koyama$^\flat$ \hspace{\authorhspace} Yuki Mitsufuji$^\flat$}\\
$^\natural$RWTH Aachen University, Germany \hspace{\authorhspace}
$^\sharp$Sony Europe B.V., Stuttgart, Germany \\
$^\flat$Sony Group Corporation, Tokyo, Japan\\
{\tt\small johannes.imort@rwth-aachen.de, giorgio.fabbro@sony.com}
}
\def\authorname{J. Imort, G. Fabbro, M. A. Martínez-Ramírez, S. Uhlich, Y. Koyama, and Y. Mitsufuji}

\usepackage[bookmarks=false,pdfauthor={\authorname},pdfsubject={\papersubject}]{hyperref}

\begin{document}

\maketitle
\begin{abstract}
Given the recent advances in music source separation and automatic mixing, removing audio effects in music tracks is a meaningful step toward developing an automated remixing system. This paper focuses on removing distortion audio effects applied to guitar tracks in music production. We explore whether effect removal can be solved by neural networks designed for source separation and audio effect modeling.

Our approach proves particularly effective for effects that mix the processed and clean signals. The models achieve better quality and significantly faster inference compared to state-of-the-art solutions based on sparse optimization. We demonstrate that the models are suitable not only for declipping but also for other types of distortion effects. 
By discussing the results, we stress the usefulness of multiple evaluation metrics to assess different aspects of reconstruction in distortion effect removal.
\end{abstract}
\section{Introduction}\label{sec:introduction}

\label{sec:intro}
With the emergence of musical recordings, audio effects have become indispensable in the music production process. They are used by musicians as a creative tool to alter the sound of their instruments, and by sound engineers to craft a balanced mix from multiple recording tracks \cite{zolzer_dafx_2011}.

For the task of mixing and automatic remixing \cite{stables_ten_2017}, the \textit{dry} (i.e., unprocessed) source tracks are required. Given the recent advances in automatic mixing \cite{martinez_ramirez_deep_2021, steinmetz_automatic_2021} and music source separation (MSS) \cite{mitsufuji_music_2022}, a system could facilitate the adjustment of a stereo mixture to the taste and preferences of the user similar to \cite{choi_amss-net_2021}.
However, when separating sources with a system trained on music stems (e.g., \verb+MUSDB18+ \cite{rafii_musdb18_2017}) the mixing process is not considered, and, hence, the output of such a system contains the \textit{wet} (i.e., processed) signal.
As nonlinear distortion is one of the most commonly used effects for electric instruments, this work focuses on musical distortion effects that are used for aesthetic means (e.g., guitar overdrive/distortion pedals) and applied in the process of mixing (e.g., tape saturation). These distortion effects result in added harmonics, intermodulation distortion, and a compressed sound \cite{wilmering_history_2020}.

This paper investigates different deep neural network (DNN) approaches regarding their applicability to the audio effect removal problem.
Our contributions can be summarized as follows:
\begin{itemize}
    \item We show that recovering the clean signal from clipped or overdriven guitar signals can be efficiently solved with neural networks designed for source separation. The models achieve high quality and fast inference in contrast to solutions based on sparse optimization.
    \item We found that the superior performance of the models evaluated on the de-overdrive task can be traced back to superimposing the overdriven signal with the clean signal. We show that the architectures are suitable not only for declipping but also for other types of distortion effects. 
    \item By discussing the results, we highlight that the metrics under evaluation prove beneficial in measuring different aspects of the reconstruction and can be advised for further investigations.
\end{itemize}

This work is organized as follows:
Sec.~\ref{sec:problem} gives a formal introduction to audio effect removal and introduces the types of distortions that were used throughout this study.
Sec.~\ref{sec:related} briefly discusses previous work on iterative and DNN-based declipping approaches. The methods under evaluation are outlined in Sec.~\ref{sec:methods}. Sec.~\ref{sec:experiments} describes the data that were used for training, reports details about the experimental setup, and presents the chosen objective evaluation metrics.
In Sec.~\ref{sec:evaluation}, we evaluate the results of the comparative study of four different neural network architectures on the task of distortion removal in guitar signals for the SoX overdrive implementation.
Then, we compare the same architectures on the declipping task to one state-of-the-art declipping algorithm using guitar signals as well as generic music signals. 
Lastly, Sec.~\ref{sec:conclusion} gives a conclusion and presents an outlook for future work.
Audio examples are available online at \href{https://joimort.github.io/distortionremoval/}{\textit{joimort.github.io/distortionremoval/}}.

\section{Problem Formulation}
\label{sec:problem}
We introduce audio effect removal as the task of recovering the original discrete audio signal $\mathbf{x} \in \mathbb{R}^n$ from the processed discrete signal $\mathbf{y} \in \mathbb{R}^n$, which is obtained by applying the possibly nonlinear and time-varying function $f$ to the signal $\mathbf{x}$:
\begin{equation}
\begin{split}
    \mathbf{y} & = g(\mathbf{x}) = \alpha f(\mathbf{x}) + (1-\alpha) \mathbf{x},
\end{split}
\end{equation}
with $\alpha \in \left[0, 1\right]$ denoting the weight of the wet signal, and $g$ the summation function of the dry and wet signal. The goal is to find an estimation of the original signal $\mathbf{\hat x}$ by estimating the inverse function $\mathbf{\hat x} = \hat{g}^{-1}(\mathbf{y})$.

Generally, distortion effects clip the input signal and can be divided into systems that apply hard-clipping or soft-clipping. 
\begin{figure}[t]
\centering
\includegraphics[width=\linewidth]{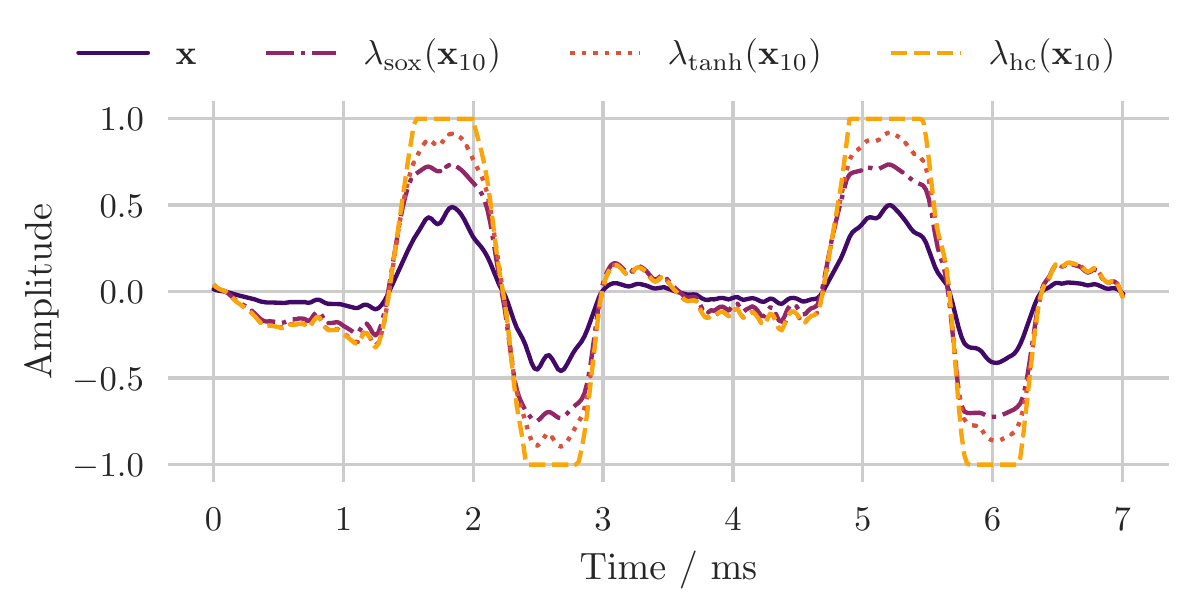}
\caption{Example of different distortion types applied to a guitar signal. The input signal $x(k)$ is amplified before being modified by a wave-shaper function.}
\label{fig:dist_types_time}
\end{figure}
As this work focuses on both types of distortion, we introduce the following generic formulation of a wave-shaper that maps the amplified signal $\mathbf{x}_{\gamma} = 10^{\frac{\gamma}{20}} \mathbf{x}$ to a fixed range:
\begin{equation}
    f(\mathbf{x}) = \lambda(\mathbf{x}_{\gamma}) \quad \textrm{with} \quad \lambda : \mathbb{R} \rightarrow [-\theta_\mathrm{c}, \theta_\mathrm{c}].
    \label{eq:waveshape}
\end{equation}
Here, $\gamma$ denotes the gain in decibels, $\lambda$ the arbitrary wave-shaper function, and $\theta_\mathrm{c}$ the fixed clipping threshold. For the case of hard-clipping, $\lambda$ is defined as:
\begin{equation}
    \lambda_{\mathrm{hc}}(x_{\gamma,k}) =
    \begin{cases}
        x_{\gamma,k}, & \text{if } |x_{\gamma,k}| \leq \theta_\mathrm{c} \\
        \theta_\mathrm{c} \mathrm{sgn} (x_{\gamma,k}) & \text{otherwise},
    \end{cases}
    \label{eq:hc}
\end{equation}
with $\mathrm{sgn}$ denoting the sign function, and $x_{\gamma,k}$ the $k$-th time sample of the amplified signal. Fig.~\ref{fig:dist_types_time} highlights the difference between different distortion types. Hard-clipping cuts off the amplitude when it exceeds a defined threshold (as typical for saturation in digital signal processing). Soft-clipping (e.g., $\lambda_\mathrm{tanh}(x_{\gamma,k})\!=\!\mathrm{tanh}(x_{\gamma,k})$) gradually applies a smooth transition before reaching a fully saturated state (as typical for saturation in analog amplifiers). 

Modeling the characteristics of distortion pedals in reality is more complex: \cite{martinez_ramirez_deep_2020} provides an overview on different methods and discusses DNN-based approaches. In order to simplify the problem for our investigation, we focus on wave-shaping. The overdrive algorithm of the audio editing software SoX \cite{bagwell_sound_2015} serves as an example of soft-clipping ($\lambda_\mathrm{sox}$) but, in contrast to (\ref{eq:waveshape}), it is dependent on previous samples. Furthermore, it blends the wet signal with the dry one ($\alpha < 1$).

\section{RELATED WORK}
\label{sec:related}
To the best of our knowledge, there has been no previous research on distortion audio effect removal. Therefore, this section outlines the most relevant iterative and DNN-based declipping approaches since declipping is a special case of distortion audio effect removal.

\subsection{Iterative Declipping Methods}
\label{sec:iterative_declipping}
Previous research on approaches to declipping has focused mainly on unsupervised algorithms that recover the signal under the assumption of a generic regularization such as signal sparsity \cite{zaviska_survey_2021}. Usually, these approaches target the 
hard-clipping case only (see (\ref{eq:hc})). While early approaches were based on auto-regressive models,
(e.g., \cite{janssen_adaptive_1986})
recent state-of-the-art methods evolved by combining ideas from inverse problems and sparse regularization \cite{gaultier_sparsity-based_2021}.

Recently, \cite{zaviska_survey_2021} and \cite{gaultier_sparsity-based_2021} discussed popular declipping algorithms. One of the current state-of-the-art methods is A-SPADE, which will serve as a baseline for this study. The algorithm is briefly introduced in Sec.~\ref{sec:methods}.

\subsection{DNN-Based Declipping}
\label{sec:dnn_declipping}
In contrast to iterative algorithms, to date, there are only few contributions comprising supervised DNNs.

Kashani \textit{et al.} \cite{kashani_image_2019} introduced a declipping method based on the U-Net architecture \cite{ronneberger_u-net_2015}. It operates on magnitude spectrograms while the waveform of the output is obtained by reusing the phase information from the distorted input signal. The system is trained and evaluated on pairs of hard-clipped and clean speech samples.

Mack and Habets \cite{mack_declipping_2019} proposed an architecture comprising a BLSTM-based deep filtering method that works on complex spectrograms and hence also considers phase information. Similar to \cite{kashani_image_2019}, they train the system on speech data only. Unlike any other approach, they not only investigate the system on the hard-clipping case but also on the soft-clipping case.

Tanaka \textit{et al.} recently proposed APPLADE \cite{tanaka_applade_2022}, a declipping method that takes advantage of the sparse optimization techniques described above together with deep learning. Accordingly, they embed a DNN in the iterative algorithm to enhance the thresholding operation. They report a slightly higher performance than previous algorithms, better robustness to mismatches between training and test data, and faster inference.

\section{METHODS}
\label{sec:methods}
This section describes four neural network architectures that we selected from the literature and evaluated on the distortion removal task. 
We approach the distortion effect removal problem from the perspective of filtering the added harmonics and intermodulation distortion, similar to \cite{mack_declipping_2019}. Therefore, we include one model from the domain of audio effect modeling and three architectures originally proposed for music source separation.\footnote{Two of the methods under evaluation, Demucs and Wave-U-Net, do not explicitly filter the signals in the audio domain, rather they perform a nonlinear mapping. Nevertheless, they were both successfully employed for MSS, which is a filtering problem.} 
For the latter models, instead of multiple output channels for different sources, we use only one output channel and consider them as general audio-to-audio transformation architectures.

\textbf{CRAFx} was proposed as a system for modeling time-varying audio effects with a neural network \cite{martinez_ramirez_general-purpose_2019, martinez_ramirez_deep_2020}. The end-to-end model operates on the signal in the time domain and is divided into an adaptive front-end (encoder), a bi-directional long-short-term-memory (BLSTM)-based structure that applies the modeled effect in the latent space, and a synthesis back-end (decoder).
In contrast to the other DNN-based models presented in this section, this architecture employs architectural priors (e.g., learnable nonlinear activations) in the context of audio effects.

\textbf{Open-Unmix (UMX)} was introduced as a reference implementation for music source separation \cite{stoter_open-unmix_2019}. The architecture is based on the BLSTM model from \cite{uhlich_improving_2017} and uses magnitude spectrograms as input features.
The essential element of Open-Unmix is its three-layer BLSTM network that enables to learn both long- and short-time dependencies \cite{hochreiter_long_1997}.
An element-wise multiplication of the input spectrograms with the estimated masks yields the final output. 
Commonly, spectrogram-based source separation models are compared with the oracle performance of an ideal ratio mask (IRM) that is defined as the ratio between the reference and the test spectrogram \cite{narayanan_ideal_2013} in decibels. For reconstruction, the phase of the input signal is used. The model was adapted for a sampling rate of $f_s=\SI{16}{\kilo\hertz}$.
We include this model in our evaluation as a standard frequency domain MSS model that relates to the BLSTM-based declipping model from \cite{mack_declipping_2019} (cf. Sec.~\ref{sec:dnn_declipping}).

\textbf{Wave-U-Net} was proposed as one of the first end-to-end approaches for music source separation based on time domain signals \cite{stoller_wave-u-net_2018}. Hence, it incorporates not only the magnitude but also the phase of music signals. It adapts the U-Net architecture \cite{ronneberger_u-net_2015} to one-dimensional audio signals. 
We decreased the models' number of learnable parameters from $17$M to approximately $1$M by reducing the number of layers from $12$ to $8$ resulting in a reduced receptive field, as we experienced overfitting for our datasets.
Since the model was successfully employed not only for source separation but automatic mixing as well \cite{martinez_ramirez_deep_2021}, we assume a general suitability for audio-to-audio transformation tasks. Moreover, the model is highly related to the U-Net-based declipping model from \cite{kashani_image_2019} (cf. Sec.~\ref{sec:dnn_declipping}).

\textbf{Demucs} was initially designed to be an end-to-end model for music source separation in the time domain \cite{defossez_music_2021}. While it builds on the Wave-U-Net model, it introduces several improvements to the architecture. The model comprises a convolutional encoder, a BLSTM structure, and a convolutional decoder.
As for Wave-U-Net, we decreased the number of trainable parameters from $66$M to $1$M by reducing the number of blocks from $12$ to $6$ resulting in a reduced receptive field.

\textbf{A-SPADE} was introduced as a sparsity-based declipping algorithm that outperforms previous similar approaches \cite{kitic_sparsity_2015}. 
For each time frame of the clipped signal $\mathbf{y}$, it approximates a solution of the following problem:
\begin{equation}
    \min_{\mathbf{x, z}} ||\mathbf{z}||_0 \mkern9mu \textrm{s.t.} \mkern9mu \mathbf{x} \in \Gamma(\mathbf{y}) \mkern9mu \textrm{and} \mkern9mu ||\mathcal{F}( \mathbf{x}) - \mathbf{z}||_2 \leq \epsilon,
\end{equation}
where $\mathbf{z}$ denotes the unknown discrete Fourier coefficients of each time frame and $\mathcal{F}$ the Fourier transform operator. $\Gamma$ is defined as the feasible space of solutions (i.e., clipping consistency constraint).
We included the algorithm in the evaluation of the declipping task as a baseline that delivers state-of-the-art performance \cite{zaviska_survey_2021, gaultier_sparsity-based_2021}.

\section{EXPERIMENTS}
\label{sec:experiments}
Our experiments focus on the following three scenarios:
\textbf{a)} We conducted experiments on guitar recordings that were processed using the overdrive algorithm of the audio editing software SoX \cite{bagwell_sound_2015} (CEG-OD).
\textbf{b)} We performed the same experiments as in the previous scenario while processing the same clean guitar recordings with hard-clipping (CEG-HC).
\textbf{c)} We tested the systems on an extensive dataset comprising various hard-clipped sounds (SignalTrain-HC) to evaluate their performance against a popular declipping algorithm when the models are trained given an increase in the amount and variety of data.

\subsection{Data}
\label{sec:data}
The models were trained on two different datasets to assess the distortion audio effect removal capabilities. The audio signals from a dataset that contains a single instrument class (e.g., electric guitar) exhibit consistent signal statistics. In order to restrict the statistics that need to be modeled in a first step, we chose to concentrate on dry guitar samples as target data.

Since a large-scale, polyphonic dataset from clean electric guitar sounds is, to the best of our knowledge, not available\footnote{A publicly available guitar dataset for the recognition of audio effects exists (IDMT-SMT-Audio-Effects \cite{stein_automatic_2010}). However, the dataset contains primarily homogeneous monophonic sounds and therefore, we choose to use the CEG dataset instead.}, we used an internal dataset, which we refer to as CEG (Clean Electric Guitar) dataset. The monaural data were gathered from various sources, mainly commercial audio loop packages and recordings of solo guitar, and has a duration of \SI{1.68}{\hour}. All signals were re-sampled to a common sampling rate of \SI{16}{\kilo\hertz} in order to speed up convergence during training.
To create the input dataset CEG-OD, the overdrive algorithm of SoX was applied to the data using five uniformly sampled gain levels in the range of $\gamma \in [20, 50]$\si{\decibel}. Likewise, the input dataset for the hard-clipping task, CEG-HC, was created using hard-clipping (see (\ref{eq:hc})) with gain levels from the same distribution. Both datasets have a total length of \SI{8.4}{\hour}.

Although CEG represents a good source of data for our experiments due to its specificity of timbre, it remains a limited resource in terms of size and variety. Before attempting to train a system to handle recordings in real environments (e.g., a commercial song), we need to investigate how the current models at our disposal handle the availability of more and diverse training data. For this purpose, we also performed experiments on the SignalTrain dataset, which consists of more than \SI{24}{\hour} of music and randomly-generated test signals \cite{hawley_signaltrain_2019}.
By applying hard-clipping to these clean data using a uniformly-sampled input SDR value in the range $\mathrm{SDR_{inp}} \in [1, 20]\si{\decibel}$, we created SignalTrain-HC. During evaluation, we applied each input SDR in the set $\mathrm{SDR_{inp}} \in \{1, 3, 5, 7, 10, 15, 20\}\si{\decibel}$ to each sample in the test set.

Each dataset was split into non-overlapping subsets for training (80\%), validation (10\%) and testing (10\%). We evaluated the models on the test split.

\subsection{Experimental Setup}
\label{sec:setup}
During the supervised training procedure, Adam \cite{kingma_adam_2015} was used as optimizer with initial learning rates according to the model's original implementations. The learning rate was reduced by a factor of 10 after 150 epochs of no decrease in the validation loss. All models were optimized using the source-to-distortion ratio (SDR) between their full output sequences $\mathbf{\hat{x}}$ and the respective target sequences $\mathbf{x}$ in each batch $\mathcal{B}$ of $N$ elements (not to be confused with the definition in the \verb+BSS_eval+ toolkit \cite{vincent_performance_2006}):
\begin{equation}
    \mathcal{L}_\mathcal{B}(\mathbf{x}, \hat{\mathbf{x}}) = \frac{1}{N} \sum_{i \in \mathcal{B}} 10 \log_{10} \left( ||\mathbf{x}_i||^2 / ||\mathbf{x}_i - \hat{\mathbf{x}_i}||^2 \right).
\end{equation}
We stopped all trainings after 1000 epochs.
All models processed audio sequences that are randomly extracted from each clip in the dataset; the length of the extracted sequences is equal to \SI{2}{\second} (\SI{2.3}{\second} for CRAFx due to its architecture). We used a batch size of 16 for all experiments.

\subsection{Objective Metrics}
\label{sec:metrics}
In speech enhancement and source separation, the ubiquitous measure to estimate the quality of a system is the SDR.
However, applying (\ref{eq:waveshape}) to a signal does not retain its energy. Therefore, we follow the approach of \cite{roux_sdr_2019}:
the \textbf{scale-invariant SDR (SI-SDR)} is obtained by rescaling the target signal $s$ such that the residual $s-\hat{s}$ is orthogonal to $s$ by using the optimal scaling factor $\hat{s}^T s / ||s||^2$.

An evaluation exclusively based on the objective similarity of the signals does not necessarily imply a correlation with human perception \cite{cano_evaluation_2016, emiya_subjective_2011}. Accordingly, we observed that the SI-SDR scores occasionally disagreed with our qualitative evaluation. Therefore, we also considered three metrics based on human perception.

The \textbf{perceptual evaluation of audio quality (PEAQ)} \cite{thiede_peaq_2000} is a widely used perceptual metric
\cite{zaviska_survey_2021, gaultier_sparsity-based_2021, lattner_stochastic_2021, you_perceptual-based_2010} that measures the amount of degradation between two audio signals.
The output of PEAQ is an Overall Difference Grade (ODG), which can reach values between $0$ (\textit{imperceptible impairment}) and $-4$ (\textit{very annoying impairment}).
Even though PEAQ is used in declipping and audio restoration studies \cite{zaviska_survey_2021, gaultier_sparsity-based_2021, lattner_stochastic_2021}, it was developed for audio codecs.

The \textbf{R-nonlin metric} \cite{tan_predicting_2004}, in contrast, was developed specifically for detecting nonlinear distortions and, like PEAQ, considers the human auditory system.
\Rnonlin{} is defined between 0 (\textit{high distortion}) and 1 (\textit{no distortion}).

The \textbf{Fréchet audio distance (FAD)} was recently proposed as a reference-free evaluation metric for music enhancement algorithms. It has shown to correlate more with human perception than the SDR \cite{kilgour_frechet_2019}.  
In order to obtain the FAD, the embedding statistics of both the whole clean and distorted test set are generated using a VGGish model \cite{hershey_cnn_2017}. The FAD is calculated based on the Fréchet distance between two multivariate Gaussians computed from both the test and the reference embeddings. \cite{dowson_frechet_1982}.

\section{RESULTS}
\label{sec:evaluation}
\begin{figure}
	\includegraphics[width=\linewidth]{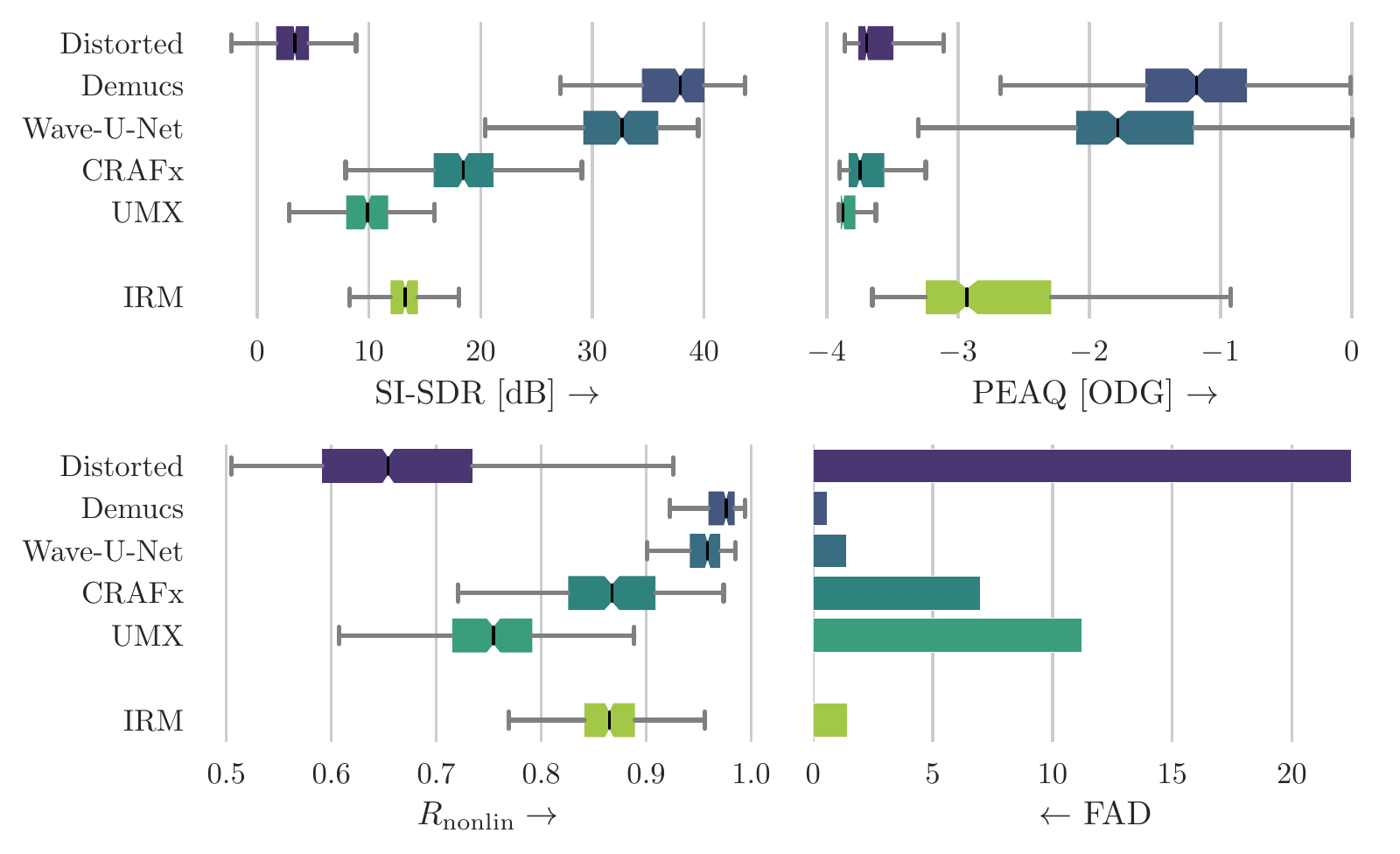}
	\caption{Box plot of scores for the CEG-OD dataset. The boxes show the first and third quartile of the data while the median is indicated with a line in the box. Higher scores indicate superior performance except for FAD. The results suggest that U-Net-based models performing convolutions in the time domain are the most promising approach to solving the task of overdrive removal.}
	\label{fig:res_CEG_overdrive}
\end{figure}
In this section, we provide the results of the experiments introduced in the previous section.
\subsection{De-Overdrive (CEG-OD)}
Fig.~\ref{fig:res_CEG_overdrive} shows the results of the models that remove overdrive from guitar tracks.

Firstly, in SI-SDR, both Demucs and Wave-U-Net perform exceptionally well and even outperform the ideal-ratio-mask by more than \SI{24}{\decibel}. While CRAFx yields considerably worse performance, it also surpasses the IRM oracle. UMX is the least performing of all the models in our comparison. It should be noted that for UMX, a model that operates on magnitude spectrograms, the IRM represents its upper limit in performance.

Similar results are obtained with PEAQ, \Rnonlin{} and FAD: while Demucs and Wave-U-Net yield the best scores and surpass the IRM, the ones for CRAFx and UMX are considerably worse.
It seems that directly processing the signals in the time domain using a U-Net-based architecture represents the most promising approach for the removal of the overdrive effect.
\subsection{Declipping (CEG-HC)}
\begin{figure}
\includegraphics[width=\linewidth]{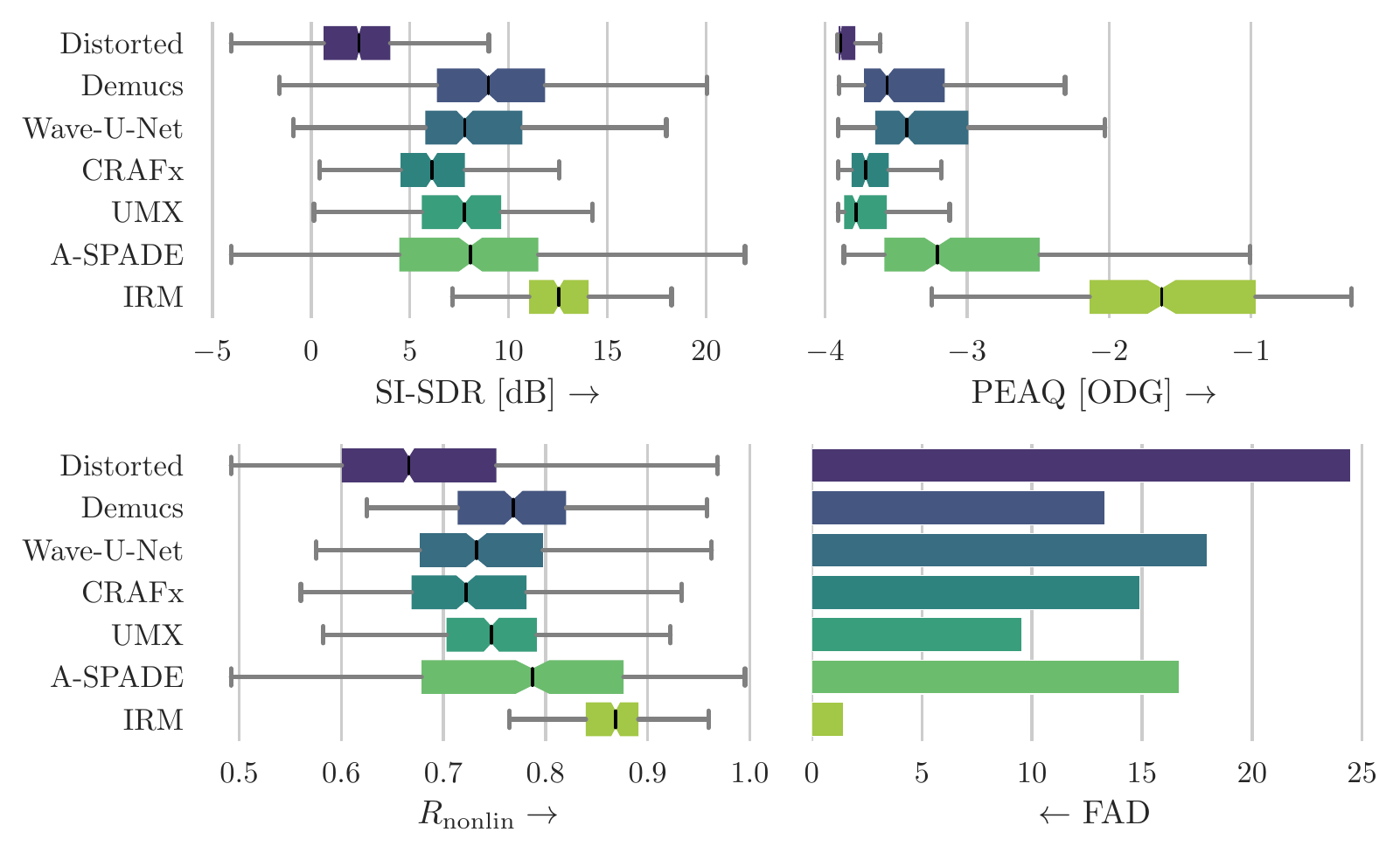}
\caption{Box plot of scores for the CEG-HC dataset.
It is difficult to clearly determine the best model, although Demucs represents a good compromise for all the metrics.
}
\label{fig:res_CEG_hardclipping}
\end{figure}
Fig.~\ref{fig:res_CEG_hardclipping} shows the results on the task of declipping on guitar recordings.
Generally, we experienced a drop in the scores: now the models have been trained on declipping, which is an ill-posed problem, as missing parts of the signal need to be reconstructed.
Additionally, we report scores for our declipping baseline, A-SPADE.

Despite the general performance drop, Demucs surpasses the results of A-SPADE in terms of SI-SDR by almost \SI{1}{\decibel}. 
While UMX and Wave-U-Net yield similar performances, CRAFx is the method with lowest scores.

Regarding PEAQ, no method surpasses the IRM and the A-SPADE algorithm. As before, Demucs and Wave-U-Net have similar performance: while PEAQ slightly favors the latter, SI-SDR favors the former. In contrast to the previous results, CRAFx has no significant advantage over UMX. 

While Demucs achieves the best score for \Rnonlin{} among the neural models, it does not surpass A-SPADE since the difference in their score is marginal. Interestingly, UMX achieves better results than CRAFx and Wave-U-Net. As the \Rnonlin{} metric was explicitly designed to detect nonlinear distortions, we conclude that UMX' outputs contain fewer nonlinear distortions than the outputs of CRAFx and Wave-U-Net. 

Surprisingly, when computing the FAD, UMX outperforms all other methods, including A-SPADE. This might be accounted to the fact that the FAD is based on the mel spectrum, whereas UMX optimizes the magnitude spectrogram. While Demucs and CRAFx outperform A-SPADE in FAD as well, the score for Wave-U-Net is slightly worse. The FAD for the IRM is relatively small because the difference between the reference and test embeddings from the \mbox{VGGish} model can be traced back to the masking operation and quantization.
Demucs seems to constitute a good compromise regarding performance since it yields first- or second-best results for all metrics. 

\subsection{Declipping (SignalTrain-HC)}
\begin{figure}
	\includegraphics[width=\linewidth]{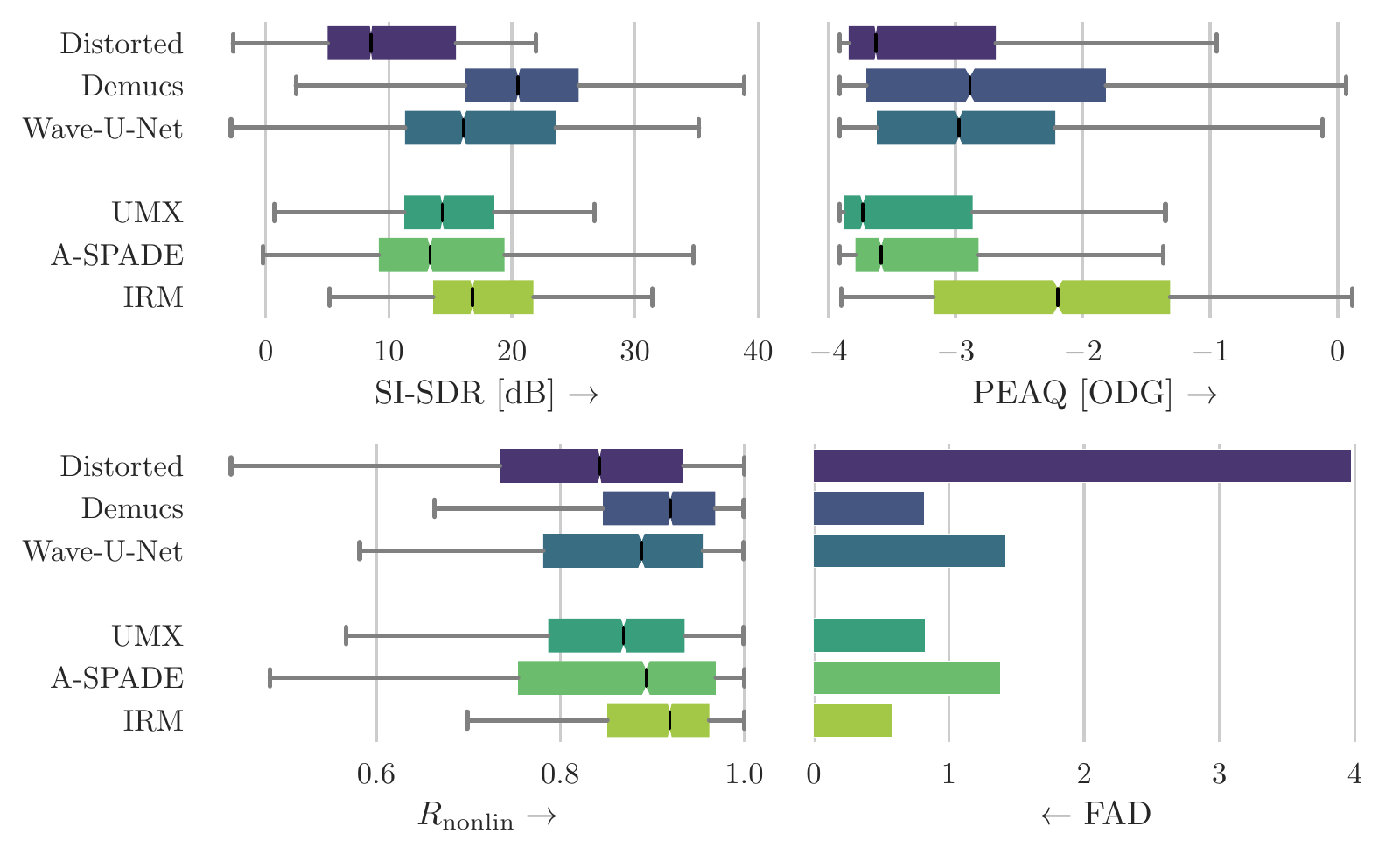}
	\caption{Box plot of scores for the SignalTrain-HC dataset. Demucs can be regarded as the best model regarding the median score across all metrics.}
	\label{fig:res_signaltrain_hardclipping}
\end{figure}
Fig.~\ref{fig:res_signaltrain_hardclipping} shows the results for declipping SignalTrain-HC. Due to the lower gains that are used to prepare the data (see Sec.~\ref{sec:data}), the overall performance seems to be superior but cannot be directly compared to the previous results. Because of its considerably worse performance in the previous task, CRAFx was left out of the evaluation. 

Demucs surpasses all other models in SI-SDR, including IRM and A-SPADE. Moreover, Wave-U-Net and UMX both surpass A-SPADE but not the IRM. 
PEAQ gives a similar ranking of the methods: only UMX cannot reach the A-SPADE baseline. None of the neural methods surpasses the IRM. In terms of \Rnonlin, the same ranking is obtained, with Demucs surpassing, Wave-U-Net reaching, and UMX just missing A-SPADE.
The FAD highlights that UMX and Demucs deliver comparable performance, outperforming the baseline, but not surpassing the IRM.

When only looking at SI-SDR or PEAQ, we notice the superiority of the time domain models. Future research should investigate whether the waveform in the time domain is the best input representation for the task, compared to, e.g., the real and imaginary part of a spectrogram \cite{choi_investigating_2020} or both the waveform and the spectrogram \cite{defossez_hybrid_2021}.
Ultimately, Demucs can be considered the best model in our experiments for the task of declipping on SignalTrain.

\subsection{Discussion}
\label{sec:discussion}
\begin{figure}[t]
\centering
\includegraphics[width=\linewidth]{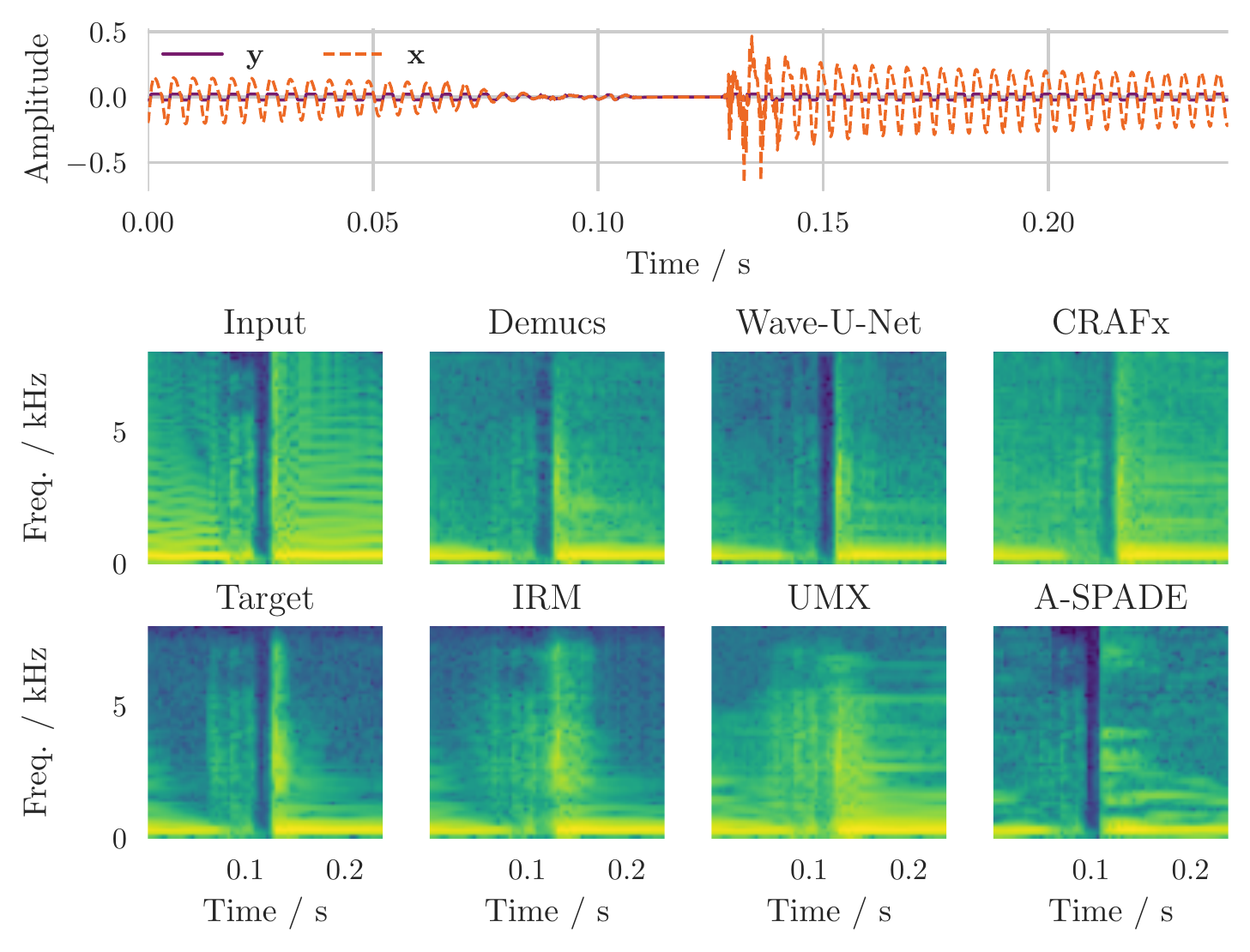}
\caption{Spectrograms of a clipped guitar signal ($\gamma\!=\!\SI{45}{\decibel}$), its target signal, and the output of each method. The respective time domain input and target signal are shown at the top
. While all methods significantly reduce the harmonics, it can be observed that IRM and UMX smear the transients. A-SPADE relies on the strongest bins in the clipped spectrum, which leads to tonal artifacts.}
\label{fig:spectrogram_comparison}
\end{figure}
\textbf{Qualitative Evaluation} \quad
Although the abundance of evaluation metrics in the literature has the potential to analyze the results in very detailed ways, it does not always aid the judgement of which method is best given the individual context.
Often, different applications have different requirements concerning the sound quality and can afford some types of distortions or artifacts to be left in the signal.
Therefore, we report some qualitative considerations that need to be taken into account concerning our results, with the aim of finding some descriptive patterns among them. Fig.~\ref{fig:spectrogram_comparison} shows spectrograms of a hard-clipped guitar signal and the outputs of all models under evaluation.

We found that the characteristics of the artifacts that each model produces are consistent, independent from the task it is applied to. 
Nevertheless, for the time domain-based models, the artifacts are less prominent in the de-overdrive task. Here, Demucs often produces outputs that are virtually indistinguishable from the original signal.

Generally, Demucs is the model that most often produces high quality results. Especially for inputs with a low amount of distortion, it can reconstruct the original sound without any perceptual artifact. 
Wave-U-Net behaves similarly, although it often cannot reach the same quality.

UMX generally removes the distortion characteristics very well at the cost of strong phasing artifacts: Despite the absence of input distortions, the spectral features of the output are not necessarily consistent with the ones of the target. This is most likely due to UMX re-using the phase of the distorted input to reconstruct the signal in the time domain and the frame-wise processing.
Moreover, Fig.~\ref{fig:spectrogram_comparison} highlights that the transients are smeared by re-using the phase of the degraded signal.

CRAFx does not suffer from phasing artifacts, but occasionally leaves part of the distortion features in the output. In some cases, the model fails to reconstruct the onset of some notes, penalizing the listening experience.

Finally, A-SPADE is the model that exhibits the strongest and most frequent artifacts, especially for strongly clipped signals. Although it considers phase information by working in the complex frequency domain, it leads to non-optimal solutions.
Nevertheless, distortion features (like those left by Demucs) or transient smearing (left by UMX) do not occur.

\textbf{Influence of the Dry Signal} \quad
We have observed a substantial difference in performance between models that need to remove overdrive (CEG-OD dataset) and models that need to remove hard-clipping (CEG-HC dataset).
The superior performance of the models trained and tested on CEG-OD can be mainly traced back to the presence of the non-distorted signal in the overdrive output and not to the soft-clipping character of the specific overdrive implementation. We verified this hypothesis by training Wave-U-Net on hard-clipped data superimposed with the clean signal (even when the amplitude of the clean signal is considerably low). We obtained results similar to those in the de-overdrive task (median results without superposition: $\textrm{SI-SDR}=\SI{7.2}{\decibel}$, with superposition: $\textrm{SI-SDR}=\SI{34.4}{\decibel}$). While the performance drop is present for all metrics, it is less pronounced for UMX which seems not to utilize the additional information in the signal.

\textbf{Inference Speed} \quad
The benefits of Demucs in the context of declipping go beyond the quality of its outputs: using a neural approach also has advantages regarding inference speed. Inference with A-SPADE is comparably slow (real-time factor on CPU $\times\!\mathrm{RT} \in [4.2, 27.3]$ depending on \SDRinp{} \cite{gaultier_sparsity-based_2021}), being an iterative approach that requires a computation of the Fourier transform and its inverse at each iteration. Demucs ($\times\!\mathrm{RT}=0.072$), Wave-U-Net ($\times\!\mathrm{RT}=0.113$) and UMX ($\times\!\mathrm{RT}=0.026$) instead, allow for fast inference on the CPU and even surpass real-time constraints independently on \SDRinp{} without sacrificing the quality of the results.

\textbf{Evaluation Metrics} \quad
The results highlight how our evaluation metrics focus on different aspects of the reconstructions: While SI-SDR measures differences between two audio signals in the time domain, PEAQ focuses on the perceptual quality without differentiating between degradation related to nonlinear distortions and artifacts/quality. In contrast, \Rnonlin{} specifically highlights nonlinear distortions that have not been removed. Finally, FAD focuses primarily on the degradation that is observable in the mel spectrum.
Hence, each metric proves beneficial in measuring specific aspects in the analysis of audio effect removal systems and can be advised for further investigations.

\section{CONCLUSION}
\label{sec:conclusion}
We showed that recovering the clean signal from clipped or overdriven audio signals can efficiently solved with neural networks designed for source separation.
We found that Demucs achieves high quality according to the chosen evaluation metrics, especially when the distortion algorithm to be removed blends the distorted sound with the original one. This outcome highlights the potential of the proposed approach for other audio effects that mix dry and wet signals (e.g., parallel compression, reverberation, delay, modulation effects).

Moreover, we showed that Demucs, Wave-U-Net, and UMX outperform one state-of-the-art declipping method on our test data. This outcome is promising, considering that the dataset to train such a system is potentially much larger than the size of the dataset we used.
By discussing the results, we stressed the usefulness of multiple evaluation metrics suitable to assess distortion removal systems. 

Future work should include gathering more clean electric guitar data and generating a dataset using high-quality distortion emulations, which is required to improve generalization on real-world data. Furthermore, the knowledge from sparsity-based declipping algorithms could yield a valuable prior for declipping through DNNs.

\bibliography{Distortion_Audio_Effects_arXiv_V1.1.bib}

\end{document}